%% file: Final_version.tex
\documentclass{aa}

\usepackage[T1]{fontenc}
\usepackage{ae,aecompl}
\usepackage{graphicx}	
\usepackage{amsmath}	
\usepackage{amssymb}	
\usepackage{txfonts}
\usepackage{float}
\usepackage{verbatim}
\usepackage{natbib}
\usepackage{textcomp}

\begin{document}
    \title{A survey for occultation astrometry of Main Belt:\\
      expected astrometric performances}
    \titlerunning{Occultation astrometry in the Main Belt}

   \author{J. F. Ferreira\inst{1,2}, 
          P. Tanga\inst{1}, P. Machado\inst{2} and E. Corsaro\inst{3}
          }

   \institute{Université Côte d'Azur, Observatoire de la Côte d'Azur, CNRS, Laboratoire Lagrange, Bd de l'Observatoire, CS 34229, 06304 Nice cedex 4, France \\
         \and
             Instituto de Astrofísica e Ciências do Espaço (IA), Universidade de Lisboa, Tapada da Ajuda - Edifício Leste - 2º Piso
             1349-018 Lisboa, Portugal \\
         \and
             INAF–Osservatorio Astrofisico di Catania, via S. Sofia 78, 95123 Catania, Italy
\\
           \email{jferreira@oca.eu}
             }

   \date{Received April 17\textsuperscript{th} 2020; accepted June 30\textsuperscript{th}, 2020}
    
    \abstract
{Occultations of stars by asteroids are an efficient method to study the properties of minor bodies, and can be exploited as tools to derive very precise asteroid astrometry relative to the target star. With the availability of stellar astrometry thanks to the ESA mission Gaia, the frequency of good predictions and the quality of the astrometry have been strongly enhanced. }
{Our goal is to evaluate the astrometric performance of a systematic exploitation of stellar occultations, with a homogeneous data set and a given instrument setup. As a reference instrument, we adopt the example of a robotic 50 cm telescope, which is under construction at the Observatoire de la C\^ote d'Azur. We focus in particular on single-chord occultations.}
{We created a data set of simulated light curves, that are modelled by a Bayesian approach. To build the final statistics, we considered a list of predicted events over a long time span, and stellar astrometry from Gaia data release 2.}
{We derive an acceptable range of observability of the events, with clear indications of the expected errors in terms of timing uncertainties. By converting the distribution of such errors to astrometric uncertainties, we show that the precision on a single chord can reach levels equivalent to the performance of Gaia (sub-milli-arcseconds). The errors on the asteroid position are dominated by the uncertainty on the position of the occultation chord with respect to the barycentre of the object.}
{The limiting factor in the use of occultation astrometry is not the light curve uncertainty, but our knowledge of the asteroid\textquotesingle s shape and size. This conclusion is valid in a wide range of flux drops and magnitudes of the occulted star. The currently increasing knowledge of the shape, spin properties, and size, must be used to mitigate this source of error.}
    
    \keywords{
    Occultations -- minor planets, asteroids: general -- astrometry -- techniques: photometric -- methods: numerical}
    
    \maketitle
    
    \section{Introduction}
    
    Transit and occultation phenomena have been exploited for decades as powerful tools to infer the properties of the occulting or the occulted object. In the case of asteroids occulting a star, the main results are size and shape measurements, which are used to study specific targets \citep{durech} or to calibrate other size-determination approaches \citep{shev}.
    
   For Centaurs and trans-Neptunian Objects (TNOs), stellar occultations offer the best opportunities to derive precious physical properties. Remarkable achievements include the discovery of a ring system around (10199) Chariklo \citep{chariklo}, the size measurement of Eris \citep{eris} and other TNOs, the determination of an upper threshold for the atmospheric density of Makemake \citep{makemake}, and the detection of variations in the atmosphere of Pluto \citep{pluto, pluto2}, to mention a few. 
    
    So far, the main challenge of occultation observations resides in the accuracy of the predictions, as the uncertainty on an asteroid ephemeris usually exceeds the size of the occultation track. A first improvement in the late 1990s, brought by the Hipparcos astrometric mission \citep{perryman}, has allowed observers to obtain more reliable predictions for asteroids with a diameter $>$50~km, occulting a sample of $\sim 10^5$ stars in the Hipparcos catalogue with the best-known astrometric parameters \citep{colas, dunham}.
    
    However, the Gaia mission by the European Space Agency (ESA) is completely changing the landscape by bringing a substantial enhancement of the prediction efficiency, thanks to the sub-mas (milli-arcseconds), precision astrometry of both target stars and asteroids.
    
    The pre-release of star positions and proper motions appearing in the second data release (DR2) \citep{gaia2, gaia3, gaia4} has demonstrated the reality of the expected improvement, with successful, accurate predictions of the occultation path for several objects, including Pluto and other TNOs such as the New Horizons mission target (486958) Arrokoth \citep{Buie_2018, Porter_2018, Ortiz_2019, Buie_2020}. 
    
    \citet{spoto} have tested the use of observed occultations as astrometric measurements, which have an accuracy comparable to that of the target star. Such an approach, if systematically applied, can give ground based observers access to ultra-precise astrometry at the level of Gaia, even after the mission has concluded its operations, with the uncertainty on the stellar proper motions as the only degrading factor of the measurement precision. 
    
    We recall here the importance of collecting accurate measurements of asteroid positions, at better than $\sim$10~mas accuracy, as this is the only way to solve some fundamental issues of asteroid dynamics, intimately linked to our knowledge about the evolution of the Solar System. Very important in this respect is the measurement of the Yarkovsky diurnal effect, which is relevant to the delivery of asteroids in the Earth space environment, and more in general to understanding the evolution of the asteroid belt \citep{vokro_2015}. We also mention the determination of asteroid masses, as a very small number of them are unambiguously known at an acceptable accuracy \citep{scheeres_2015}.
    
    The astrometric data set based on charge-coupled device (CCD) observations, available from the Minor Planet Center ($>$200 millions observations), has an average accuracy of $\sim$400~mas \citep{desmars2013}. Astrometry by the Hipparcos mission or by stellar occultations exhibits better residuals (10-150 mas) but they represent $\sim$0.5\% of the total. Radar ranging measurements have a comparable performance but, again, on a very limited sample mostly composed of near Earth asteroids (NEA) \citep{ostro_2002, farnocchia_2015}. Due to the rapid fading of the signal intensity with distance, only a few large main belt objects have been observed with this technique. In summary, the total number of measurements with an accuracy of $\sim$10 mas is a tiny fraction, represented by $\sim$10,000 entries. 
    
    In this context, the Gaia DR2 asteroid data set alone, with two million astrometric positions for asteroids at $\sim$mas accuracy, already represents a spectacular improvement. However, the time interval over which the observations have been obtained is rather short. A joint use of DR2 with pre-Gaia data is then necessary to detect secular effects. 
    
    Even after the last data release by Gaia covering the entire operational phase (currently accepted for extension to 2022), the task of securing very accurate astrometric measurements over longer time spans will be necessary to improve Yarkovsky detections. In particular, detecting Yarkovsky directly in the main belt will be of the highest importance, to use it as a ''clock'' for better constraining the age of the collisions that created asteroid families \citep{spoto_2015}. As the semi-major axis change rate due to the Yarkovsky effect is stronger for smaller asteroids (going as $D^{-1}$) and weaker at increasing heliocentric distance (as $r^{-2}$), to reach its detection in the main belt, asteroids with a diameter of D$\lesssim$10~km must be targeted.
    
    The forthcoming Legacy Survey of Space and Time (LSST) by the Vera C. Rubin observatory, will certainly contribute to this effort through the discovery and follow up astrometry of a large number of small asteroids, both in the NEA and main belt population \citep{ivezic_2019}. It will operate between 2023 and 2033, with two more years added to the mission for final data processing. Its contribution will impact all targets fainter than magnitude 16 (saturation limit of the telescope), however, the single-epoch astrometric accuracy will be limited to 10-20 mas. 

    The only technique capable of obtaining, at least in principle, an astrometry close to Gaia quality, is the observation of stellar occultations on a systematic basis, on asteroids smaller than those considered in predictions up to now. This approach will clearly disclose the possibility to extend in time the collection of the best possible astrometry and push the limits to detect subtle dynamical effects.
    
    The expansion of the number of usable stellar targets (the Gaia sample is approximately complete at all magnitudes G$<$20.5) and the strong quality improvement for $\sim$3$\times 10^5$ asteroids are at the origin of the revolution triggered by Gaia. While asteroid orbits improved by Gaia astrometry are gradually published and optimised, the number of potentially observable events increases by orders of magnitude  \citep{tanga1}. 
    
    This change of perspective clearly suggests that fixed, specialised telescopes can become very efficient and observe several occultations per night. Such a systematic, massive exploitation clearly calls for automated systems to observe, process and store the collected data.
    
    In the framework of a project named UniversCity, the Observatoire de la C\^ote d'Azur (OCA) is building a 50~cm robotic telescope that will start to operate at the Calern observing station in 2020. 

    A full description of the instrument and its operation is beyond the scope of this article. Rather, our main goal is to assess the possible astrometric performance of occultation astrometry for a large sample of main belt asteroids, when a single observing site is considered. As we discuss further below, this is difficult to obtain by exploiting the existing data set of observations. For this reason, we present an independent assessment, corresponding to constant conditions of observations, a homogeneous set of data representing occultation light curves, a given  instrument, and an identical data reduction approach. Our results demonstrate the performance that can be reached under controlled, constant, and nearly ideal conditions. They can be rescaled to different instruments and observational conditions. We use them to determine which events can be considered to be observable (i.e. capable of providing usable data) and the astrometric accuracy that can be obtained from them. 

We stress here that our focus is on the massive exploitation of a statistically significant amount of data obtained by a single fixed instrument, capable of exploiting the expanded domain of exploitable predictions enabled by Gaia. The original aspect of this work is that such an approach is clearly different from the common networked observation, in which several telescopes (usually portable ones) are deployed to cover an event by specific, high priority objects (binary asteroids, TNOs, mission targets). We will further detail the  general properties of occultation observations available in the archives and their associated astrometry in the next section. Section 3 illustrates the simulations that we use to explore a homogeneous set of light curves, which are then fitted to a simple model as explained in Sect. 4. In Sect. 5 we show the results of the fit, from which a performance in the astrometry is derived (Sect. 6). The main conclusions and limitations of our approach are summarized in Sect. 7.
    
\section{Properties of currently available asteroid astrometry}

Stellar occultations of asteroids are being observed by networks of dedicated amateurs, with a few coordinators who compute and update predictions, collect the results, and process the data to derive information on the occulting asteroid shape, size, and astrometry. Free software is also available for the different steps of this process. Occultations present a variety of difficulties (mostly a function of target star magnitude, maximum duration, and flux drop), and are observed with a corresponding variety of instruments, ranging from small photocamera objectives to large, professional telescopes.

Since the 1990s, video and CCD cameras have become the standard tool for these observations. An accurate time reference is provided by the one-pulse-per-second (1 pps) of a global positioning system (GPS) receiver, but sometimes less accurate timing is still adopted, as provided by internal computer clocks (synchronized - or not - by external devices or by network time protocol servers). Although not completely error-free, the use of the 1 pps signal of GPS is an affordable approach that can easily reach an $\sim$1~ms accuracy when electronically coupled to the shutter signal of the acquisition camera.
  
The timing accuracy is very relevant to our research as the epoch of disappearance and reappearance of the occulted stars are the fundamental quantities that are obtained from the observation. As the apparent motion of the asteroid with respect to the target star can be predicted with high precision, the timings can be converted to an accurate astrometric position in the direction of the asteroid motion itself. We will call this measurement the "along track" (AT) position or astrometry. 

In the direction perpendicular to the asteroid motion (i.e. "across track", or AC), the uncertainty of the astrometry can vary, depending upon the knowledge of the physical properties (size, shape) of the asteroid. Even the ideal case of a spherical occulter can produce a large uncertainty on the AC position when its size is not well constrained, unless several occultation chords are observed from different telescope sites. 

The frequent case of elongated or irregular asteroids can introduce even more uncertainty into the determination of both the AT and AC position when single occultation chords are available, unless the shape and its orientation are known.

\begin{figure}
    \includegraphics[width=\hsize]{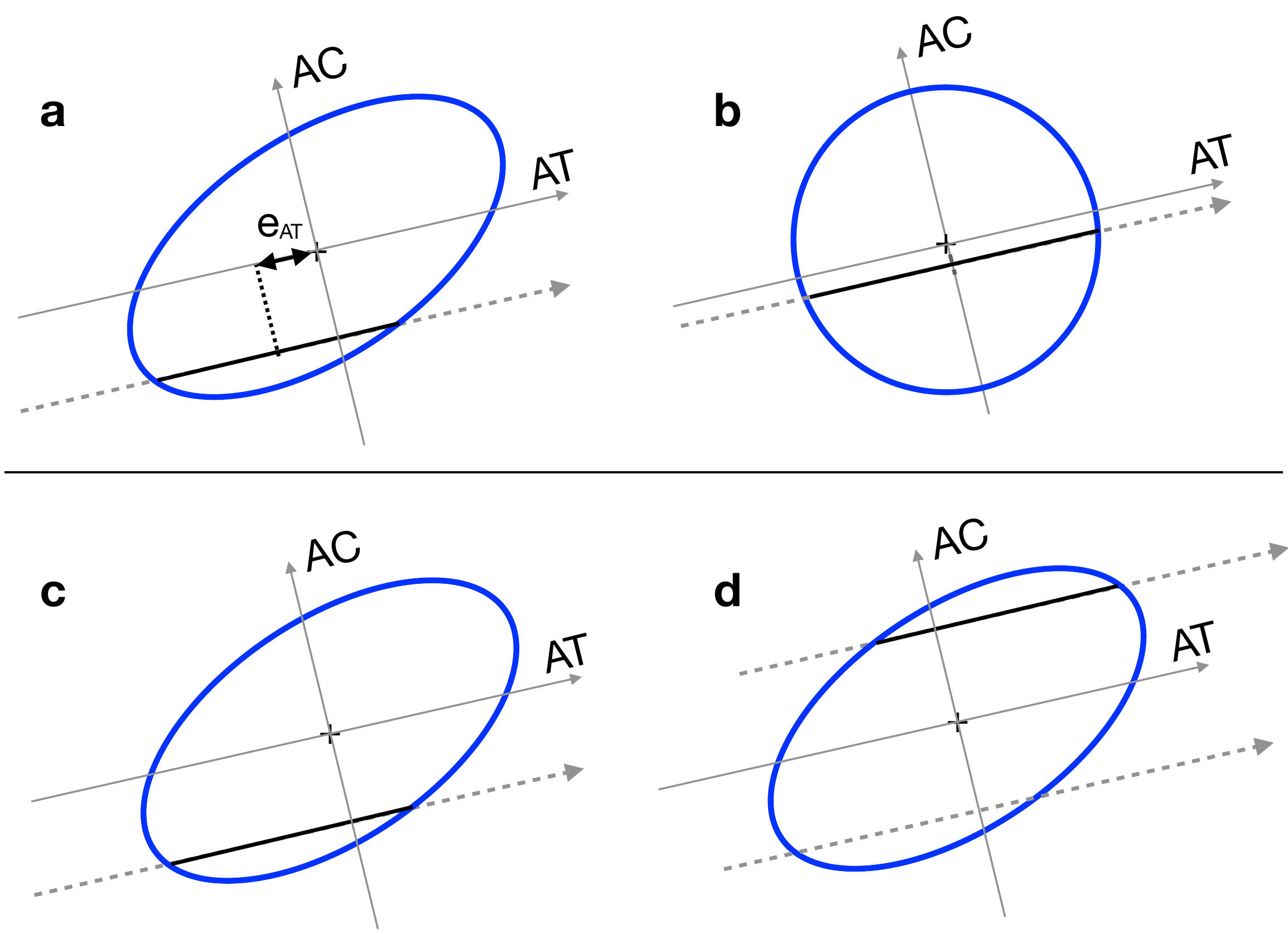}
    \caption{idealised scheme of the uncertainties involved in the astrometry derived from single chord observations of stellar occultations. We adopt as an example a very elongated ellipsoidal asteroid, projected on the fundamental plane in which the asteroid is at rest and the occulted star moves along the dashed arrows. The AC and AT axes represent the across and along--track directions, respectively. In (a) the observed chord duration (black segment) is the same as for the surface-equivalent sphere in (b). The mid-chord point corresponds to the barycentre position in (b), but not in (a), where the error on the derived astrometry e$_{AT}$ cannot be estimated without a precise knowledge of the shape. Points (c) and (d) represent the ambiguity on the AC position of the observed chord, which can be on opposite sides of the barycentre.}
    \label{F:ellipses}
    \end{figure}
Single-chord events, suffering from such limitations, are a majority, as they represent about 60$\%$ of the whole data set of occultations \citep{herald_2019}. As we will deal with the capabilities of a single, isolated telescope, it is important to take such shortcomings into account. 
    
Figure~\ref{F:ellipses} illustrates the errors involved in the determination of the AC and AT astrometry for a non-spherical shape, which can be mitigated (but not completely eliminated for single chords) if the shape is known.

The most common error is in fact the AC ambiguity, due to the fact that a given chord duration can occur on both sides of the occultation path (Fig.~\ref{F:ellipses}, panels (c) and (d). In this case, the probability distribution of the AC position is strongly bi-modal.

Extremely elongated or irregular shapes can increase such errors, which in the worst case reach $\sim$half the largest extension of the asteroid silhouette. For this reason, their relevance decreases with the object size and is lower than $\sim$10 mas for objects of 20-30~km diameter in the main belt.    

To assess the expected performance of occultation astrometry for a dedicated robotic telescope, it would be tempting to use the current data set of observed events derived from the observations that are available through the Planetary Data System base \citep{herald_2019} or the Minor Planet Center. This data set directly provides the final astrometry, represented by the relative position between the star and the asteroid at the moment of the closest appulse, as seen from the geocentre. Its derivation is strongly model dependent, as it can involve shape fitting to the observed chords, as well as complex, approximated estimations of the uncertainties. 

For these reasons, we prefer to take control of the whole process and proceed to an evaluation of the telescope performances based on simulated signals, whose properties are calibrated on real, measured quantities and known CCD camera specifications. Most importantly, this approach allows us to extend the investigation to poorly known regimes of ''difficult'' events (i.e. short durations, small flux drop) that are not adequately represented among past recordings. Occultation observers usually favour high--probability events that have good predictions and are not at the limit of their instruments. This approach introduces a clear bias that we intend to overcome, as a robotic telescope can more easily be programmed to spend time on less probable or difficult events.

\section{Occultation light curves: Simulations}
    
We aim to study a reliable method to retrieve the parameters of a large number of occultation light curves, capable of operating on a variety of signal properties. To achieve this goal, we simulated a set of occultation signals, with different durations of the occultation and signal--to--noise ratios (S/N). 
 
 We simulated stellar occultation light curves as if they had been observed by imaging at visible wavelengths and as if the photometry of the occulted star, plus a reference, had been acquired by aperture photometry after the usual corrections (bias and dark frame subtraction) and normalisation (flat field). 
    
    We also consider that the asteroid flux, right before and after the event, falls within the aperture of the target star, such that at no time are the two sources resolved. Our reference for the out-of-occultation flux will thus be a luminosity level corresponding to the combination of the two sources. This is the usual situation, as their images are completely merged in the proximity of the event epoch.
    
    Also, we consider that the Fresnel diffraction \citep{roques, pass} is negligible, as the typical Fresnel length for a main belt asteroid is \textasciitilde300~m, much smaller than the asteroid size range that we consider in this article ($>$5~km). At the frame rate adopted in the following sections (0.1~s), time resolution is not sufficient to reveal the diffraction pattern drifting at 10-20~km/s over a ground-based telescope.
    
    We can also completely neglect the apparent star size in this context, as for all stars with $V_*>11$ it is well below 1~mas. We have verified this assumption by using the stellar radii and parallaxes published in Gaia DR2.  Although these have been obtained without corrections for the interstellar reddening, they are expected to be accurate enough for an estimate at $\sim$20$\%$ level \citep{andrae_2018}. In the data set, we find a maximum apparent diameter of $\sim$0.2~mas at $V_*=11$, which is too small to become a serious concern. 
    
\begin{figure}
    \includegraphics[width=\hsize]{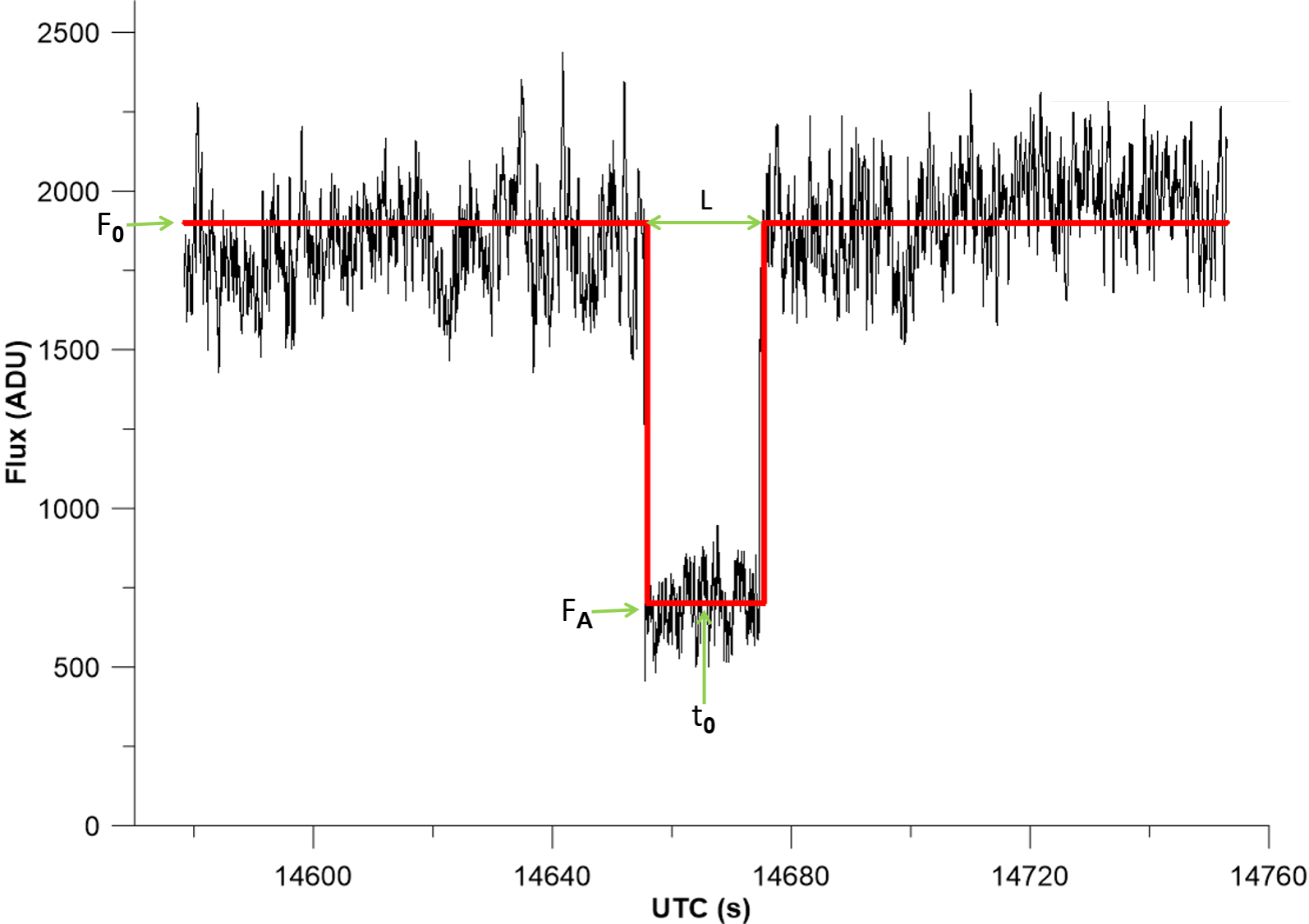}
    \caption{Example of a reduced light curve with the resulting model fit (as described in Sect.~\ref{S:regression}). Asteroid (41) Daphne occulted a $V\sim$9.3 star on March 2, 2017. From the fit (red curve), the measured occultation duration was 18.8 $\pm$ 0.2 seconds, and the drop ~63\% of the original flux, or 1.08 $\pm$ 0.01 magnitudes. Data points of the flux outside the occultation extend beyond the plotted range on both sides, which contributed to the calculation of the combined flux.}
    \label{F:Daphne}
    \end{figure}
        
    We adopt a conservative transmission efficiency (telescope reflections, correcting optics, and camera included) of 0.3. Some observations performed at the site of the UniversCity telescope with a $\sim$1 m instrument, with an Andor iXON 888 camera and a Johnson-Cousin standard R filter, have allowed the determination of a typical background value used for our simulations of $N_{bg}$ = 7.400~ADU~s$^{-1}$ pixel$^{-1}$, (ADU - analog-to-digital unit) per unit square metre of the mirror surface. These are re-scaled to the size of the UniversCity telescope, by assuming the same transmission factors, as well as the same camera and filter (R).
    
    By taking into account photon noise from the sources, background, read--out noise from the Andor iXON 88 camera, and scintillation noise caused by atmospheric turbulence, we find that, for an exposure time $dt=0.1\ $s, the star's scintillation dominates up to magnitude $V_*\sim$9 and the background, read-out, and dark-current noise sources take over from $V_*>$11. A practical limit for a star at a low zenith distance turns out to be  $V_*\sim$13 at S/N$\sim$5. We model all the noise sources, including scintillation, following \citet{mary} to which we refer for details.
    
To better sample the limits of faint drops and signals (S/N$\sim$3), and to include unfiltered imaging (R filter removed) with respect to the limiting magnitude estimate above, we do not consider very bright stars (those contributing only a fraction of events in practice) and focus on the magnitude range $V_*=11-14$. For this same goal, we also select events whose ratio between duration and exposure time is in the range of two to ten. 
    
Our model for the light curves consists in a flat signal outside and inside the occultation and a sharp transition at the beginning and end of the event, and it is described (as illustrated in Fig.~\ref{F:Daphne}) by the following parameters:
    
    \begin{itemize}
    \item $F_0$: The combined, summed flux from the asteroid, $F_A$, and the target star, $F_*$, before and after the occultation. The corresponding magnitude is defined as $V_c$.
    \item The flux drop $F_0-F_A=F_*$, corresponding to a magnitude drop: $V_{drop}=2.5\ Log\left[(F_*+F_A)/F_*\right]$.  
    \item $L$: The duration of the occultation.
    \item $t_0$: The central epoch of the occultation.
    \end{itemize}
    
     The parameter ranges for our simulations are as follows:
    
    \begin{itemize}
    \item Star magnitude $V_* \in [11.5, 14.0]$.
    \item Magnitude drop $V_{drop} \in [0.3, 1.0]$.
    \item Duration $L \in [0.2, 1.0]$~s.
    \end{itemize}
    
The exposure time is considered to correspond to the sampling interval, an assumption that is valid in general for fast video cameras such as the iXON mentioned above. For this reason, hereinafter we adopt simply ''exposure time'', although a readout time must be accounted for, depending on the acquisition device.
    
The simulations also adopt a constant exposure time, $dt = 0.1$~s. The mid-point $t_0$ is assumed to fall around the centre of the simulated time span, but its exact timing is shifted with a uniform probability distribution within an interval $\pm$ $dt / 2$, to randomise the phase of the sampling with respect to the central epoch of the minimum. Discrete intervals of (0.2 magnitudes) were used as a distribution for $V_*$.  

We note that magnitude drops exceeding our chosen interval are possible when faint asteroids are involved. However, larger drops are easier to observe, so their contribution to assess the limitations of our approach is minor.

Noise is added to the signal by random sampling of a Gaussian distribution, whose variance is computed from the combination of the various sources \citep{mary}.
    
    We note that when acquiring a real signal on the sky, the flux $F_0$ can be obtained with very high accuracy before and after the occultation, provided that the sky is photometrically stable, by measuring it on a time interval (for instance, a few minutes) much longer than the occultation event itself (a few seconds). Also, multiple reference stars can be used to minimise spurious fluctuations in the measured flux. Assuming that this procedure is validated for a specific setup, the uncertainty on $F_0$ can be considered negligible.
    
We can expect that the factor dominating our capacity to retrieve the parameters of the occultation signal is related to the signature of the light drop with respect to the noise. We thus define the drop--to--noise ratio (DNR) expressed as:
    \begin{center}
    $DNR = \frac{F_0-F_A}{\sqrt{{\sigma_{(F_0)}}^2+{\sigma_{(F_A)}}^2}}$ 
    \end{center}
    
    where $\sigma_{F_0}$ and $\sigma_{F_A}$ are the noise of the signal outside and inside the occultation. Our results in the following confirm that the DNR is the main parameter characterising the ``difficulty'' of the data reduction for each light curve. 
    
\begin{figure}[ht!]
    \includegraphics[width=\hsize]{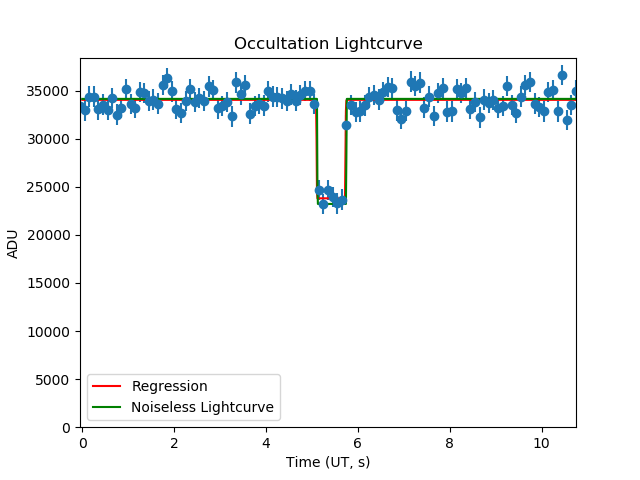}
    \includegraphics[width=\hsize]{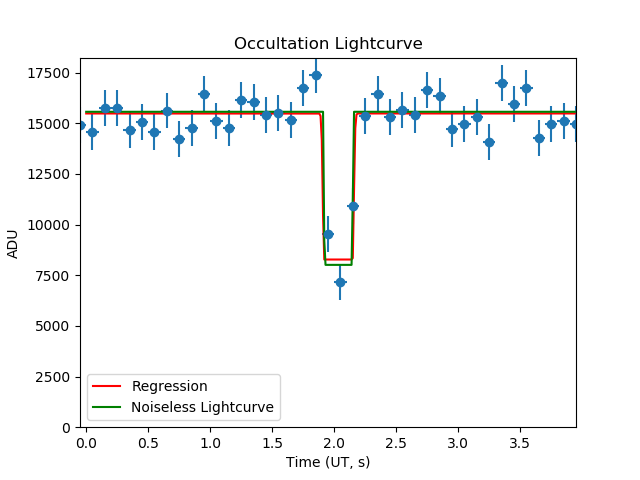}
    \caption{Examples of simulated light curves (green line), sampled with our noise model at dt$=0.1\ s$ (blue dots). The associated blue bars represent the one-sigma uncertainties derived from the noise model. Regression is obtained through the method explained in Sect. \ref{S:analysis} (red). Top panel: $V_*=12.5$, drop of 0.4 magnitudes, nine samples, DNR = 4.9. Bottom panel: $V_*=13.5$, drop of 0.7 magnitudes and two samples, DNR = 2.1.}
    \label{F:examples}
    \end{figure}

To sample the parameter ranges defined above, a total of 7920 light curves were simulated. Two examples of simulated light curves that have different minima durations and DNRs are shown in Fig.~\ref{F:examples}.

In addition, we test the presence of false positives, which should start to appear when the expected DNR is very low. Our approach is to use the same parameters for the simulations exploited above, but with the elimination of the occultation signature in the data. The processed signal is thus just the constant out-of-occultation level, $F_0$, with its noise. We still apply priors as if the occultation was present (next section).     

Eventually, we note that the discretisation of $V_*$ in the simulation parameters, will show up in the graphic representation of our results, under the form of clusters around DNR values. We are aware of this limitation, but we will see that the conclusions are statistically robust and would not change with the adoption of a more uniform distribution.

    \section{Regression procedures on the simulated data}
    \label{S:regression}
    
{To retrieve the light curve parameters from the simulated data sets described above we adopt the Bayesian inference method (BIM) by a nested sampling Monte Carlo algorithm as implemented in the \textsc{DIAMONDS}\footnote{ https://github.com/enricocorsaro/DIAMONDS} package \citep{corsaro1, corsaro2, corsaro3}.}

To apply the BIM, we implement a light curve function as:

    \begin{equation}
        \label{E:model}
        F(t) = F_0-F_A\ S(\sigma, t_0, L)
    \end{equation}

where S is a ''supergaussian'' function composed of three branches:
    
   \begin{center}
   S($\sigma$, $t_0$, $L$) $=$ \[ \left\{
    \begin{array}{ll}
          N(t, t\textsubscript{0}-L/2, \sigma) & t \leq  t\textsubscript{0}-L/2\\
          1 & t\textsubscript{0}-L/2\leq t\leq t\textsubscript{0}+L/2\\
          N(t, t\textsubscript{0}+L/2, \sigma) & t \geq  t\textsubscript{0}+L/2\\
    
    \end{array} 
    \right. \]
    \end{center}

In the equation above, $N(x, x_c, \sigma)$ is a normal distribution function of $x$, centred at $x_c$, of width $\sigma$. With the small value for $\sigma$ that we adopt (1/10 of the exposure time) the supergaussian becomes similar to a ''gate'' function, but it preserves its properties of being continuous with continuous derivatives. For our specific case of occultation events, the possibility of the BIM to exploit priors is a very relevant advantage, as the predictions of occultation events provide exploitable information on the expectations for each of the light curve parameters, in particular the expected maximum duration, flux drop and centre epoch of the event.

Each simulated light curve is described by a combination ($V_*$,$V_{drop}$, L, $t_{0}$) or equivalently in term of fluxes ($F_0-F_A$, $F_A$, L, $t_{0}$). These parameters are related to the observation of a hypothetical predicted event. In practical applications, the possible range of values that they can assume is constrained by the a priori knowledge conveyed by the prediction, which informs us about the brightness of the star and the asteroid, the maximum expected duration, and other factors. As a consequence, defining Bayesian priors to reduce our simulated light curves is equivalent to using the predicted parameters of an occultation event to reduce its observed light curve. The only difference is that in the case of a real event the probability distribution of the priors is constrained by the prediction; in the case of simulations, it is constrained by the nominal values used to produce each light curve. In both cases we need to make reasonable guesses about the width of the distribution, which we assume to be Gaussian. 

In the case of real observations, that exploit predictions, the total flux $F_0$ can be estimated from the brightness of the target star and the asteroid, by taking into account the properties of the instrument and the observing conditions. The flux drop $F_0 - F_A$ can also be estimated, since the theoretical brightness of the asteroid can be affected by a rather large error (several tenths of magnitude) due both to the absolute magnitude $H$ errors and to the viewing or illumination geometry of a possibly irregular shape.
    
The maximum duration $L_{max}$ is provided in general by the prediction, but can be affected by a large uncertainty, depending both upon the uncertainty on the object size, and on the projection of its shape at the epoch of the event.

Eventually, knowledge of the absolute occultation epoch t$_{0}$ is dominated by the uncertainty on the asteroid orbit and the star position. For typical multi-opposition main belt objects, the uncertainty on t$_{0}$ is of the order of a few seconds. In practice, the observation always extends for a much longer time before and after t$_{0}$.  
    
On the basis of these considerations, we define Gaussian priors whose averages are derived from the nominal parameters ($F_0-F_A$, $F_A$, L, $t_{0}$), and standard deviations consistent with those of the usual predictions. In particular:
       
    \begin{itemize}
    \item $\sigma_{F_0} = 0.5$ mag and $\sigma_{F_*} = 0.3$ mag. This choice is done to include the discrepancy present in the prediction of asteroid brightness, which is well known to be affected by inaccurate absolute magnitude \citep{pravec_2012}, slope parameters and shape--related effects. The star magnitude is much better known in general in the current catalogues, but variability and colour effects can be present, so we prefer to be conservative.  
    
    \item $\sigma_{L} = 0.2 \times$ the maximum duration ($L_{max}$). This parameter is affected by our knowledge of asteroid size and shape. Our choice is compatible with uncertainties for asteroids of a few kilometres in diameter in the NEOWISE survey. \citep{masiero_2011, mainzer_2011a, mainzer_2011b} This choice - while not excluding that regression from reaching a shorter duration - implicitly introduces the assumption that central chords (diametrical events) are more probable. A more sophisticated approach could be to impose a realistic probability distribution for the chord duration.
    
    \item $\sigma_{t_0} = 3$~s. For each prediction, by propagating the asteroid orbit errors and the error estimate on the star astrometry, an associated uncertainty on the central epoch can be computed. However, in the process, the contribution of astrometric biases on the orbit accuracy (especially those present in old asteroid astrometry) are easily underestimated. We adopt a single fixed value here, to mitigate this issue and avoid overly optimistic uncertainties. 
    \end{itemize} 
     
Our choices for the definition of the priors are very general. The adoption of Gaussian priors is a compromise permitting us a first exploration of the performance of the BIM on the simulations. When applied to real observations, more complex prior distributions can be used, adjusted for the distribution of the uncertainty on the asteroid ephemeris, the star astrometry and brightness. Varying sky conditions and possible timing errors can also affect the prior shape and spread. The associated difficulty of implementing the corresponding automated processing, capable of taking them into account, are outside the scope of this article, but can be included in practical applications. 

      \begin{figure*}[ht]
    \centering
    \includegraphics[width=0.45\hsize]{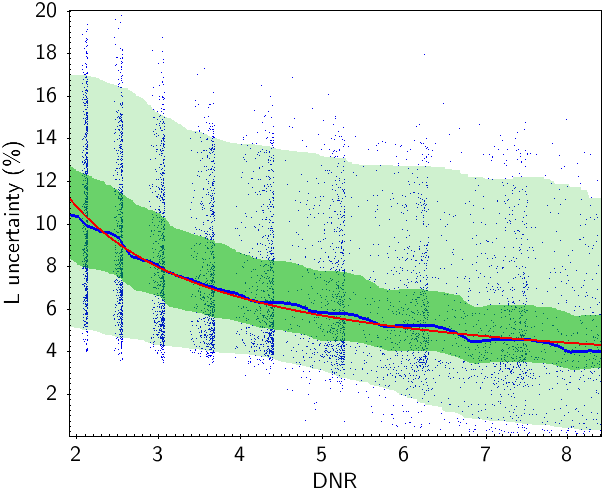}
    \includegraphics[width=0.45\hsize]{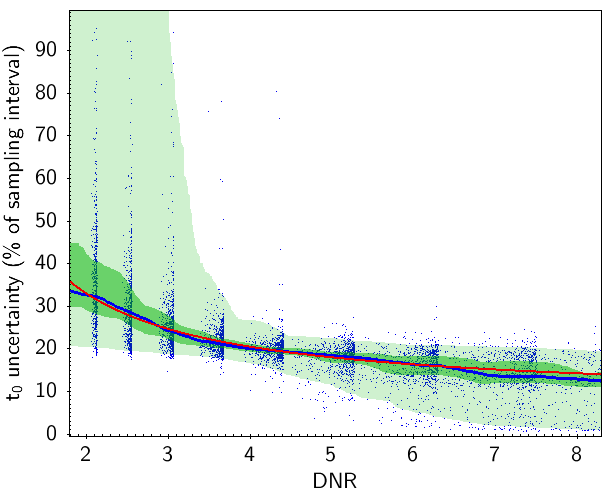}
    \caption{Uncertainty on the duration L and the centre epoch $t_0$ of the occultation, as a function of the DNR, estimated by the standard deviation of the posterior distribution, after Bayesian modelling. Each dot represents a result obtained from a simulated occultation. The continuous blue line is the smoothed average value, while the shaded areas enclose the quantiles equivalent to 1-$\sigma$ and 2-$\sigma$. The trend  $\sim$1/DNR, explained in the text, is represented by the red line. The uncertainty is expressed, respectively, as percentage of duration, and of the exposure time. The clustering of the blue data points around discrete DNR values is a consequence of the choice of a discrete $V_*$ distribution, but it does not affect the estimation of the general trend.}
    \label{F:BIM_uncertainty}
    \end{figure*}
    
\section{Results of the regression on the simulated light curves.}
\label{S:analysis}
    
    The significance of the occultation can then be evaluated by comparing the Bayesian evidence of two light curve  models: one containing the occultation signal $E_{occ}$; the other, a constant (no occultation) function $E_{flat}$.\footnote{We recall here that the Bayesian evidence (or marginal likelihood) is given by the average of the likelihood distribution over the parameter space set by the priors.} From the processing of our simulated light curve, we compute the detection probability of the occultation using the evidence ratio $C = E_{occ}/(E_{flat} + E_{occ})$. We consider that a model is strongly significant with respect to another if the Bayesian evidence of the former is at least five orders of magnitude larger than that of the latter model, yielding a $C_{crit}\sim 0.993$. We thus classify as "non significant" all results with $C<C_{crit}$ \citep{corsaro1, corsaro2}.
      
    \begin{figure}[hb!]
    \centering
   \includegraphics[width=\hsize]{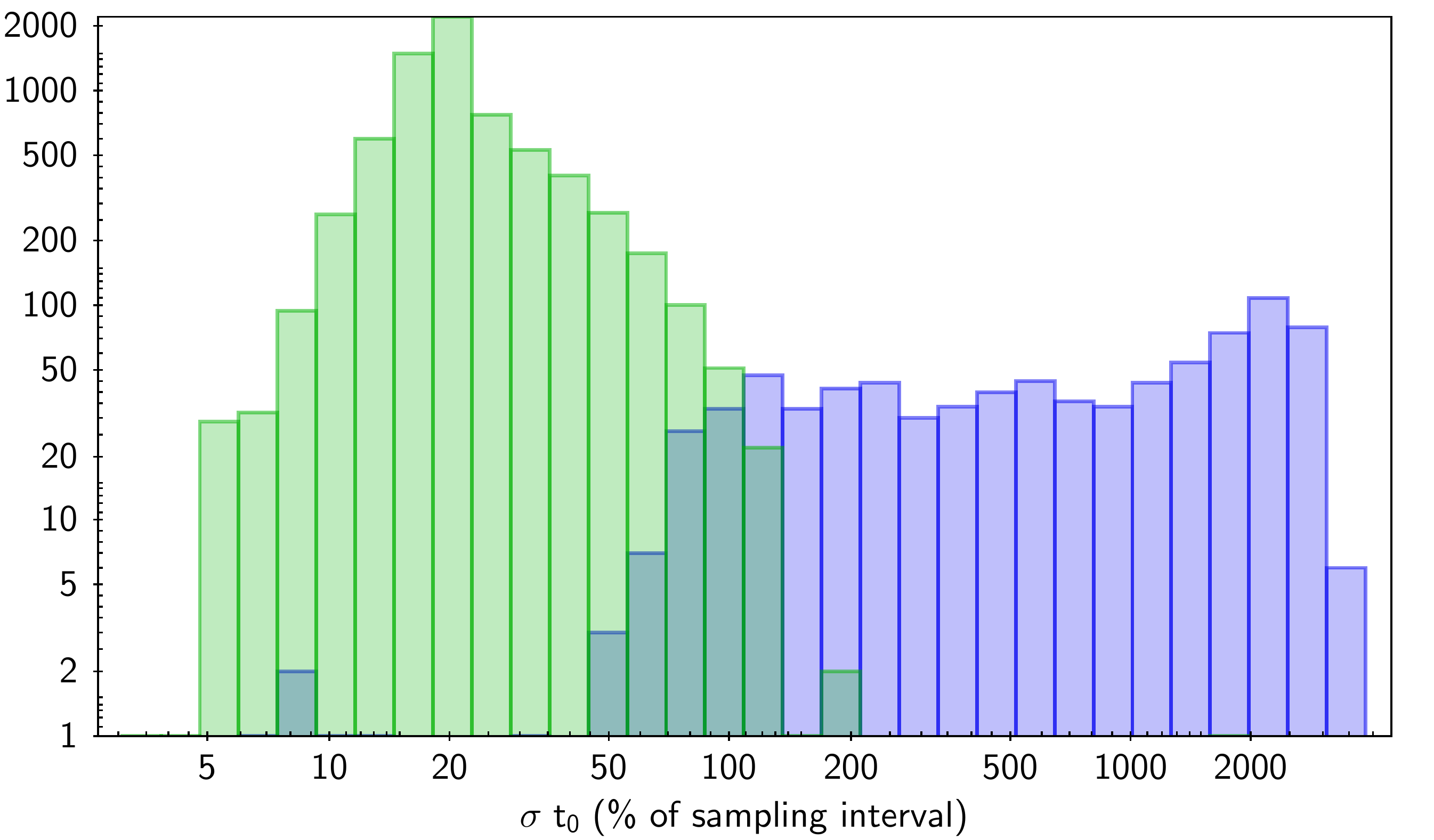}
    \caption{Distribution of the significant (left, green peak) and non-significant (right, blue peak) solutions as a function of the uncertainty on the centre epoch $t_0$, expressed as a percentage of the exposure time. At around 100$\%$ uncertainty, the frequency of the two situations is about the same.}
    \label{F:Precision_Significant}
    \end{figure}

In the following, we characterise the results obtained from the simulated light curves on the basis of the uncertainty (standard deviation) of the marginal distribution for the duration and the centre epoch ($\sigma_L$ and $\sigma_{t_0}$), which are the two fundamental parameters determining the observed chord length and the position of the asteroid relative to the star, in the direction of its motion. While the first quantity is related to the physical properties of the object (size, shape), the second directly defines the quality of the occultation astrometry, the seminal parameter that motivates our study.
    
    Hereinafter, $\sigma_L$ is expressed as a percentage of the total duration, while $\sigma_{t_0}$ is a percentage of the exposure time. As in general the occultation minimum is sampled by several intervals, we find a smaller uncertainty for a higher DNR (Fig.~\ref{F:BIM_uncertainty}). Interestingly, there is a clear trend of increased error dispersion for DNR$<$3, implying that the determination of $\sigma_{t_0}$ in limit conditions, despite a good average performance, can be unreliable. However, with the exception of such extreme situations, the average of $\sigma_{t_0}$ is well below the exposure time. The overall trend of the average uncertainties is $\sim$1/DNR, as expected for a trend described by a general S/N.
     
    The fraction of false positives and non-significant solutions are additional indicators of our capabilities of retrieving the occultation parameters. We can observe in Fig.~\ref{F:Precision_Significant} that at $\sigma_{t_0} > 100\%$ non-significant solutions start to dominate the population. However, we see that already at $\sigma_{t_0}\sim 50\%$ about 1$\%$ of the solutions are false positives. As such a validation criterion based on a threshold for uncertainties only, can still allow for a small percentage of false positives.

A more stringent criterion comes from the detection probabilities. A comparison of the distribution of the false positives with respect to the true events, shows that false positives are systematically characterised by a low detection probability, $C<C_{crit}$. For the events that pass the threshold of $C_{crit}$, the fraction of false positives becomes negligible ($\sim5\ \times\ 10^{-4}$).
    
For completeness, we also tested the same regression on simulated light curves using a different shape of prior distribution, namely uniform probabilities, which have a constant, non-zero values within a limited interval. In this case, we obtain a worse performance. Also, while the Gaussian priors always produce a final fit, the uniform priors do not converge for 6\% of the cases where the Gaussian priors are successful. 
            
    \section{Derived uncertainty on asteroid astrometry}
    
By a simple procedure, we can convert the precision that we expect on the centre epoch of the occultation chord (Fig.~\ref{F:BIM_uncertainty}) to the corresponding along-track astrometric precision.
    
    For the asteroids from the orbit database at the Minor Planet Center, we proceed by computing the predicted circumstances for occultation events from a single arbitrary site (in our case Nice observatory, MPC code 020) over a full year\footnote{By the asteroid occultation package "Linoccult", http://andyplekhanov.narod.ru/occult/occult.htm}. We consider a sample of Gaia DR2 target stars with V$<$14, and include asteroids with a diameter $>$5~km. For each predicted event, we compute the expected DNR with our telescope configuration, and the corresponding apparent angular topocentric velocity vector $\vec{v}_{a}$ of the asteroid, in the equatorial frame, projected on the sky. 
    
    The apparent velocity is the scale factor that allows us to convert the uncertainty on timing, given by the average shown in (Fig.~\ref{F:BIM_uncertainty}), to an angular uncertainty on the position of the single chord relative to the occulted star, simply by the relation $\sigma_{t_0} * \vec{v}_{a}$.
    
    \begin{figure*}[ht]
    \includegraphics[width=0.52\hsize]{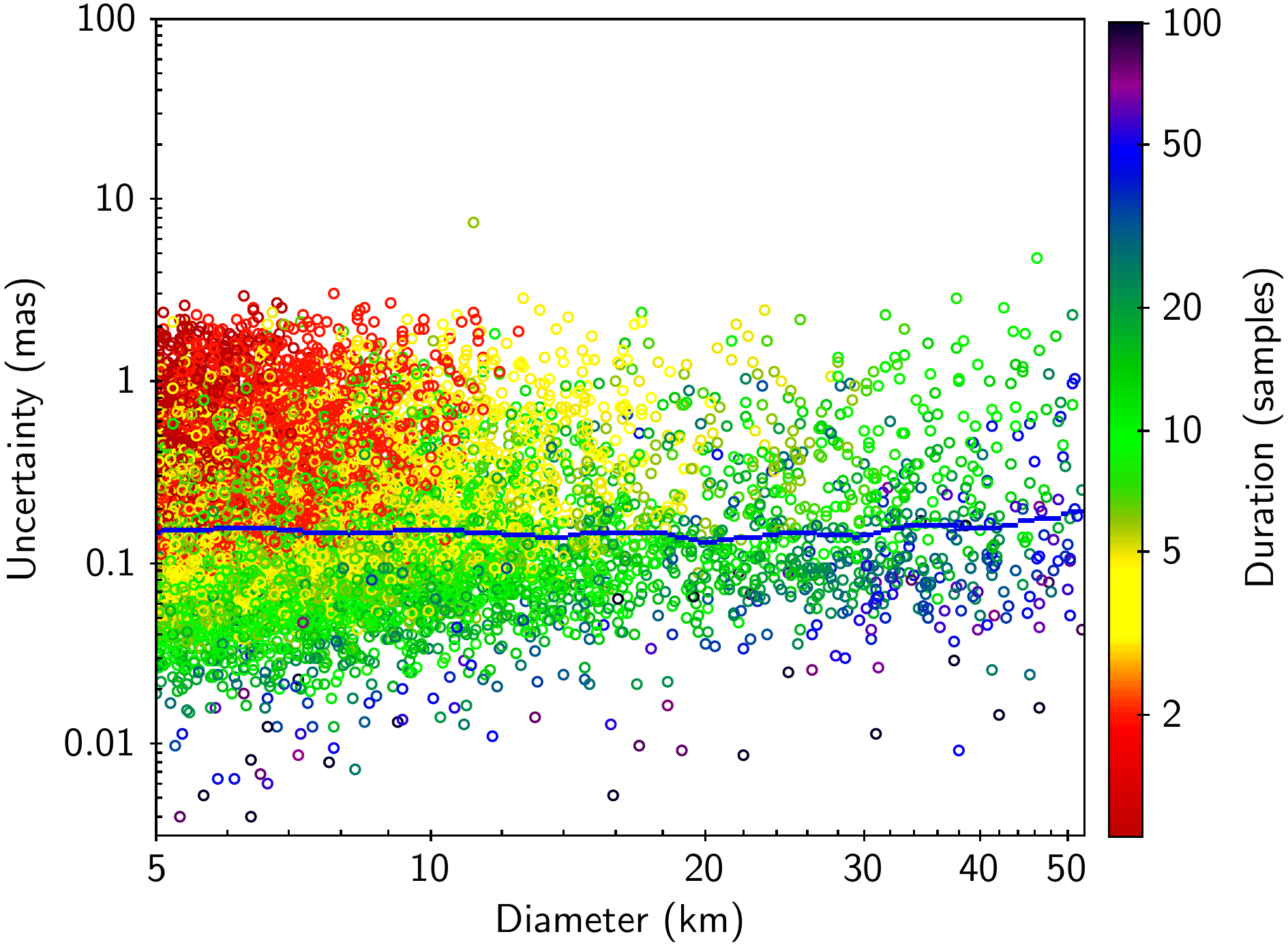}
    \includegraphics[width=0.45\hsize]{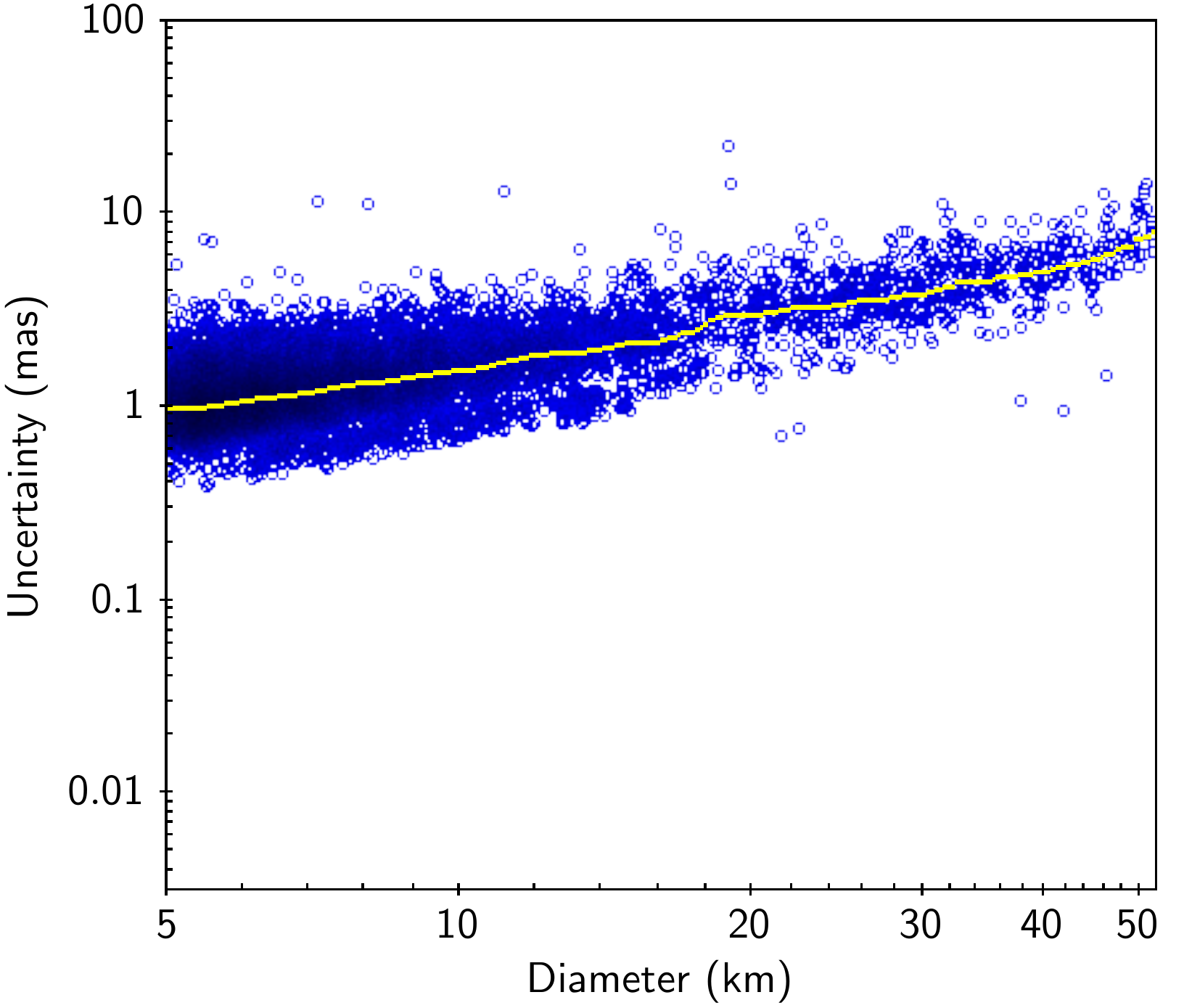}
    \caption{Expected distribution of the astrometric uncertainty on the single chord estimated by our accuracy model, applied to a large set of predictions. In the left panel, only the component of the uncertainty relative to the fit of each light curve by the BIM is taken into account. The colour corresponds to the chord length, expressed as the number of data points within the light curve minimum. The right panel plots the uncertainty including the AT error as derived from a simple model (half of the asteroid apparent radius, computed for each event). The continuous lines are the smoothed averages of the two distributions. For D$<10$~km and occultation duration $<$5 samples, the uncertainty due to asteroid size no longer dominates and the distributions overlap.} \label{F:Uncertainty_distributions}
    \end{figure*}

    \begin{figure}[hb!]
    \centering
    \includegraphics[width=1.0\hsize]{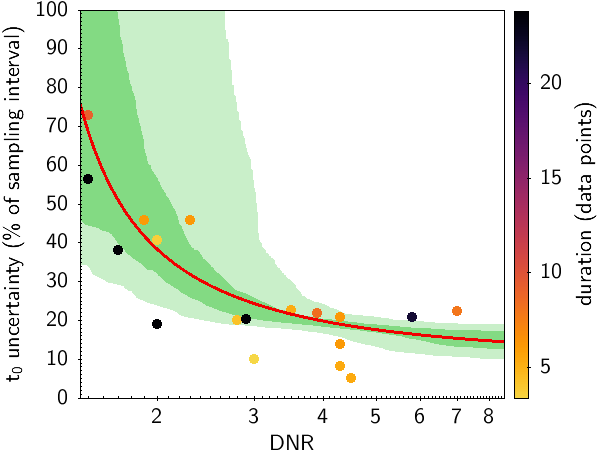}
    \caption{Drop-to-noise (DNR) ratio and uncertainty on central epoch for a number of observed events similar to those adopted for the simulation. They appear as coloured circles superposed on the general trend of Fig.~\ref{F:BIM_uncertainty} from which we reproduce only the average (red) and the one to two sigma levels. The colour of the circles is associated to the number of data points in the occultation minimum, indicated by the scale at the right of the plot.}
    \label{F:real_uncertainty}
    \end{figure}

    The resulting distribution (Fig.~\ref{F:Uncertainty_distributions}) is clearly clustered towards sub-mas values, with a peak at $\sim$0.3~mas, showing all the potential of occultation astrometry. The colour-coded distribution also shows that chords of shorter length are in general affected by a larger uncertainty.
    
However, in practical application this excellent performance has to be weighed against the fact that a single chord is a poor approximation of the asteroid centre of gravity, as discussed in Sect. 2 (Fig.~\ref{F:ellipses}). In Fig.~\ref{F:Uncertainty_distributions} we compare the uncertainty on the single chord to the uncertainty due to missing knowledge of the asteroid shape ($e_{AT}$ in Fig.~\ref{F:ellipses}). We model $e_{AT}$ as a fixed quantity of half the object apparent radius, computed at the epoch of each occultation event. We stress here the fact for exploiting the astrometry in orbit computation, both AC (across-track) and AT (along-track) error sources must be included in an adequate error model, however we focus here on the AT direction, as it is the one directly linked to the timing errors that we analyse.

The along-track uncertainty, $e_{AT}$, dominates the error on the chord timing for objects $>$15~km. At the opposite end of the size range, a non-negligible fraction of small asteroids are affected by the two error sources, fit uncertainty and $e_{AT}$, in a more balanced way, in particular when the chord length is $\le$4 sample intervals. This regime corresponds, as expected, to the rather poor sampling of the light curve minimum, which reduces the performance of the fit. By considering a typical apparent motion of a main belt asteroid (5-10 mas/s) in the inner belt (where $\sim$1~mas $\sim$ 1~km at opposition), and our time sampling of 0.1~s, the corresponding physical chord length is $\sim$1.5-3~km. This represents a practical limit, which can be overcome only by increasing the sampling frequency, if the available flux allows.

For larger objects, the overall error budget seems to be dominated by shape and size effects, increasingly for larger asteroids, and one should note that if these physical properties have been independently measured, they can be exploited to correct the astrometry derived by the occultation. Today, some thousands of shapes and rotational parameters (period, spin axis) are known, mainly from rotation light curve inversion. The largest database of asteroid sizes ($\sim$10$^5$), is produced by the cited NEOWISE survey. As the number of asteroids for which shape and size is known will increase in the future, the possibility to determine the position of the centre of mass with respect to the observed chords will be more and more common, permitting us to get closer to the limit represented by our single-chord astrometric accuracy. 
    
Eventually, an evaluation of the final accuracy will be provided by the use of single-chord occultation astrometry in the improvement of orbits, where post-fit residuals can be used to evaluate anomalies in the adopted error model and to identify the outliers. This exploration, done on the data set currently available and extending \citet{spoto} will be part of a separate article. 

Here, we limit our comparison to some real positive events, by processing their light curve with the same approach used for the simulations. We looked for a set of events observed with video or CCD cameras over the last approximately ten years, and that have both a DNR and a duration in the appropriate range for the comparison (for instance ''easy'' events with very high DNR are of little interest here). For the chosen light curves, the sampling is regular and no missing data are present. This is clearly a small sample, but despite a growing number of reports, only a fraction of observers provide the photometric data and satisfy at the same time our selection criteria.

The selected sample consists of 25 events. The DNR is estimated directly from the light curve, in the absence of the required information for more accurate modelling. Drop-to-noise ratio computation shows that 18 events fall in the range of interest to us (Table~\ref{T:realdata}).
       
We notice (Fig.~\ref{F:real_uncertainty}) that the overall trend is satisfactorily respected. A larger dispersion is present, which is probably the result of the large variety of instruments and observational conditions, certainly  much less homogeneous than in our simulated sample. We consider that within the limit of this comparison our model appears to be close enough to the reality to be usable for an approximate estimate of the performance of our general survey. The lack of a clear dependence on the number of data that sample the light curve minimum, as long as there at least two or three data points, further strengthens the role of the DNR in the characterisation of the performance.

    \section{Conclusions}
    \label{S:conclusions}
    
With the help of simulations, we show the capabilities of detecting and measuring occultation signals by a specific telescope setup in the idealised framework of homogeneous data quality, focusing in particular on the uncertainty on chord length ($\sigma_L$) and central occultation epoch ($\sigma_{t_0}$). Our statistics focus on main belt asteroids, the largest population offering both scientific challenges and the highest frequency of occultation events. 
    
We show in quantitative terms that $\sigma_{t_0}$ reaches a sub-sampling resolution, with a sensible degradation of the performance for DNR$<$3-4. In practice, when evaluating the results on real light curves for low DNRs, without knowing a priori the reality of a detection, only solutions with C$>C_{crit}$ should be selected. The resulting probability of having a false positive will then be $\sim 5 \times 10^{-4}$.

Our results depend essentially on the DNR and (to a smaller extent) on the number of samples in the occultation minimum. The performances of other instruments and acquisition setups can be evaluated by computing the aforementioned quantities.

Another factor that we do not consider here is represented by the possible systematic effects that are not included in our noise error budget. The most important are the possible presence of time correlation in the noise (Fig.~\ref{F:Daphne}) and the presence of fluctuations related to unstable transparency conditions. Also, the possible effect on the light curve of stellar multiplicity (that shows up sometimes in the form of stepped light drops) can be considered in future and need a different modelling approach.

In fitting real events, another adaptation will be needed, by changing the Gaussian probability on the duration to a distribution that is less penalizing for chords that are shorter than the maximum duration, if  the occultation is not central. In the context of our article this is not a limitation, as all our simulations include durations close (at noise level) to their nominal maximum values.

We also find that single-chord accuracy becomes relevant for a non-negligible fraction of occultation events involving asteroid of sizes D$<$15~km, for which the absolute uncertainty on the barycentre derived from a single chord becomes sensitive to size and shape. This is especially true for chords of low duration (i.e. few data points in the occultation minimum). With the improving accuracy of occultation predictions for smaller asteroids and fainter stars, an increasing fraction of observed events will fall into this domain. In particular, robotic telescopes can systematically attempt events whose probability is moderate (for instance 20-30$\%$): these will typically fall in the above range. This is also the size range of interest for obtaining new astrometry of target candidates for the detection of the Yarkovsky effect.

We stress that we do not develop here the problem of the error model in the AC direction, which is entirely size and shape dependent and should include the ambiguities illustrated in Fig.~\ref{F:ellipses}. In using occultations for orbit refinement, this component must also be carefully evaluated. A recipe for the complete error model (AC, AL) and its use for orbit refinement will be detailed in a forthcoming article.

The issue of the model dependence of occultation astrometry, involving knowledge of shape, size, and rotational properties, is mitigated by the increasing knowledge that we have of them. We mention here the fact that single occultation chords will eventually contribute to the reconstruction of shapes and sizes. In fact, even isolated events have an astrometry accurate enough to improve orbits and subsequent occultation predictions. For specific objects, the prediction can become reliable enough to justify the deployment of mobile telescope arrays, thus effectively constraining the asteroid size through multiple chords. In this respect, we can conclude that the synergy between fixed telescopes hunting for single chords and traditional, multiple sites targeting specific events with improved predictions, is a very promising and fertile development in the domain of stellar occultations.

\section*{Acknowledgements}
    
This research has been supported by the Pessoa program for science cooperation between Portugal and France, the Programme Nationale de Planetologie in France, the BQR programs of Laboratoire Lagrange and Observatoire de la C\^ote d'Azur. E.C. is funded by the European Union Horizon 2020 research and innovation program under the Marie Sklodowska-Curie grant agreement No. 664931. P.M. acknowledges project FCT 395 P-TUGA PTDC/FIS-AST/29942/2017.
    
A special thanks to Rui Agostinho (IA) and  Jean-Pierre Rivet (OCA), for their help on some critical aspects of our light curve modelling, and to the anonymous referee who helped to considerably improve the manuscript.

We acknowledge the use of the Linoccult package by A. Plekhanov to compute the predictions of asteroid occultations described in Sect. 6. For this task we made use of data from the European Space Agency (ESA) mission {\it Gaia} (\url{https://www.cosmos.esa.int/gaia}), processed by the {\it Gaia} Data Processing and Analysis Consortium (DPAC, \url{https://www.cosmos.esa.int/web/gaia/dpac/consortium}). Funding for the DPAC
has been provided by national institutions, in particular the institutions participating in the {\it Gaia} Multilateral Agreement.

This research made use of Astropy,\footnote{http://www.astropy.org} \citep{astropy:2013, astropy:2018}; matplotlib \citep{Hunter2007}; scipy \citep{scipy2019}.
    
 
    \bibliographystyle{aa} 

    \bibliography{mybib}

\section*{Appendix}
\include{tab_real}

\end{document}

%% file: tab_real.tex
\begin{table*}[ht!]
\centering
\begin{tabular}{|cl|ccc|c|c|}
\hline
Number & Name & Year & Month & Day & Observer & DNR \\
\hline
10 & Hygiea & 2015 & 5 & 14 & Dunford & 1.5 \\
65 & Cybele & 2015 & 10 & 28 & Kattendidt & 1.5 \\
25 & Phocaea & 2017 & 3 & 3 & Dunford & 1.7 \\
412 & Elisabetha & 2016 & 6 & 20 & Dunford & 1.9 \\
21 & Lutetia & 2016 & 8 & 13 & Perello & 2.0 \\
476 & Hedwig & 2019 & 2 & 12 & Conjat & 2.0 \\
580 & Selene & 2019 & 7 & 22 & Haymes & 2.3 \\
131 & Vala & 2016 & 1 & 2 & Perello & 2.8 \\
2226 & Tchaikovsky & 2019 & 1 & 17 & Conjat & 2.9 \\
306 & Unitas & 2015 & 4 & 30 & Bardecker & 3.0 \\
429 & Lotis & 2016 & 3 & 1 & Gault & 3.5 \\
479 & Caprera & 2017 & 5 & 4 & Hooper & 3.9 \\
146 & Lucina & 2016 & 11 & 30 & Frappa & 4.3 \\
117 & Lomia & 2011 & 5 & 26 & Gault & 4.3 \\
135 & Hertha & 2015 & 9 & 25 & Perello & 4.3 \\
270 & Anahita & 2017 & 3 & 26 & Hooper & 4.5 \\
3200 & Phaethon & 2019 & 10 & 15 & Tanga & 5.8 \\
221 & Eos & 2015 & 9 & 26 & Perello & 7.0 \\
\hline
\end{tabular}
\caption{Observed occultation chords chosen for comparison to our model for observation accuracy. Their photometric light curves data have been fitted by the model, with the same procedure adopted for the simulations.}
\label{T:realdata}
\end{table*}

%% file: Final_version.bbl
\begin{thebibliography}{42}
\expandafter\ifx\csname natexlab\endcsname\relax\def\natexlab#1{#1}\fi

\bibitem[{Andrae {et~al.}(2018)Andrae, Fouesneau, Creevey, Ordenovic, Mary,
  Burlacu, Chaoul, Jean-Antoine-Piccolo, Kordopatis, Korn, Lebreton, Panem,
  Pichon, Thévenin, Walmsley, \& Bailer-Jones}]{andrae_2018}
Andrae, R., Fouesneau, M., Creevey, O., {et~al.} 2018, Astronomy \&
  Astrophysics, 616, A8, publisher: EDP Sciences

\bibitem[{{Astropy Collaboration} {et~al.}(2013){Astropy Collaboration},
  {Robitaille}, {Tollerud}, {Greenfield}, {Droettboom}, {Bray}, {Aldcroft},
  {Davis}, {Ginsburg}, {Price-Whelan}, {Kerzendorf}, {Conley}, {Crighton},
  {Barbary}, {Muna}, {Ferguson}, {Grollier}, {Parikh}, {Nair}, {Unther},
  {Deil}, {Woillez}, {Conseil}, {Kramer}, {Turner}, {Singer}, {Fox}, {Weaver},
  {Zabalza}, {Edwards}, {Azalee Bostroem}, {Burke}, {Casey}, {Crawford},
  {Dencheva}, {Ely}, {Jenness}, {Labrie}, {Lim}, {Pierfederici}, {Pontzen},
  {Ptak}, {Refsdal}, {Servillat}, \& {Streicher}}]{astropy:2013}
{Astropy Collaboration}, {Robitaille}, T.~P., {Tollerud}, E.~J., {et~al.} 2013,
  \aap, 558, A33

\bibitem[{{Braga-Ribas} {et~al.}(2014){Braga-Ribas}, {Sicardy}, {Ortiz},
  {Snodgrass}, {Roques}, {Vieira-Martins}, {Camargo}, {Assafin}, {Duffard},
  {Jehin}, {Pollock}, {Leiva}, {Emilio}, {Machado}, {Colazo}, {Lellouch},
  {Skottfelt}, {Gillon}, {Ligier}, {Maquet}, {Benedetti-Rossi}, {Gomes},
  {Kervella}, {Monteiro}, {Sfair}, {El Moutamid}, {Tancredi}, {Spagnotto},
  {Maury}, {Morales}, {Gil-Hutton}, {Roland}, {Ceretta}, {Gu}, {Wang},
  {Harps{\o}e}, {Rabus}, {Manfroid}, {Opitom}, {Vanzi}, {Mehret}, {Lorenzini},
  {Schneiter}, {Melia}, {Lecacheux}, {Colas}, {Vachier}, {Widemann},
  {Almenares}, {Sandness}, {Char}, {Perez}, {Lemos}, {Martinez},
  {J{\o}rgensen}, {Dominik}, {Roig}, {Reichart}, {Lacluyze}, {Haislip},
  {Ivarsen}, {Moore}, {Frank}, \& {Lambas}}]{chariklo}
{Braga-Ribas}, F., {Sicardy}, B., {Ortiz}, J.~L., {et~al.} 2014, \nat, 508, 72

\bibitem[{{Buie} {et~al.}(2018){Buie}, {Porter}, {Verbiscer}, {Leiva},
  {Keeney}, {Tsang}, {Baratoux}, {Skrutskie}, {Colas}, {Desmars}, {Stern}, \&
  {New Horizons MU69 Occultation Team}}]{Buie_2018}
{Buie}, M., {Porter}, S.~B., {Verbiscer}, A., {et~al.} 2018, in AAS/Division
  for Planetary Sciences Meeting Abstracts \#50, AAS/Division for Planetary
  Sciences Meeting Abstracts, 509.06

\bibitem[{{Buie} {et~al.}(2020){Buie}, {Porter}, {Tamblyn}, {Terrell},
  {Parker}, {Baratoux}, {Kaire}, {Leiva}, {Verbiscer}, {Zangari}, {Colas},
  {Diop}, {Samaniego}, {Wasserman}, {Benecchi}, {Caspi}, {Gwyn}, {Kavelaars},
  {Ocampo Ur{\'\i}a}, {Rabassa}, {Skrutskie}, {Soto}, {Tanga}, {Young},
  {Stern}, {Andersen}, {Arango P{\'e}rez}, {Arredondo}, {Artola}, {B{\^a}},
  {Ballet}, {Blank}, {Bop}, {Bosh}, {Camino L{\'o}pez}, {Carter},
  {Castro-Chac{\'o}n}, {Caycedo Desprez}, {Caycedo Guerra}, {Conard},
  {Dauvergne}, {Dean}, {Dean}, {Desmars}, {Dieng}, {Bousso Dieng}, {Diouf},
  {Dorego}, {Dunham}, {Dunham}, {Durantini Luca}, {Edwards}, {Erasmus}, {Faye},
  {Faye}, {Ferrario}, {Ferrell}, {Finley}, {Fraser}, {Friedli}, {Galvez Serna},
  {Garcia-Migani}, {Genade}, {Getrost}, {Gil-Hutton}, {Gimeno}, {Golub},
  {Gonz{\'a}lez Murillo}, {Grusin}, {Gurovich}, {Hanna}, {Henn}, {Hinton},
  {Hughes}, {Josephs}, {Joya}, {Kammer}, {Keeney}, {Keller}, {Kramer},
  {Levine}, {Lisse}, {Lovell}, {Mackie}, {Makarchuk}, {Manzano}, {Mbaye},
  {Mbaye}, {Melia}, {Moreno}, {Moss}, {Ndaiye}, {Ndiaye}, {Nelson}, {Olkin},
  {Olsen}, {Ospina Moreno}, {Pasachoff}, {Pereyra}, {Person}, {Pinz{\'o}n},
  {Pulver}, {Quintero}, {Regester}, {Resnick}, {Reyes-Ruiz}, {Rolfsmeier},
  {Ruhland}, {Salmon}, {Santos-Sanz}, {Santucho}, {Sep{\'u}lveda Ni{\~n}o},
  {Sickafoose}, {Silva}, {Singer}, {Skipper}, {Slivan}, {Smith}, {Spagnotto},
  {Stephens}, {Strabala}, {Tamayo}, {Throop}, {Torres Ca{\~n}as}, {Toure},
  {Traore}, {Tsang}, {Turner}, {Vanegas}, {Venable}, {Wilson}, {Zuluaga}, \&
  {Zuluaga}}]{Buie_2020}
{Buie}, M.~W., {Porter}, S.~B., {Tamblyn}, P., {et~al.} 2020, \aj, 159, 130

\bibitem[{{Colas} {et~al.}(2011){Colas}, {Frappa}, {Lecacheux}, {Berthier},
  {Vachier}, \& {Maquet}}]{colas}
{Colas}, F., {Frappa}, E., {Lecacheux}, J., {et~al.} 2011, in EPSC-DPS Joint
  Meeting 2011, 1538

\bibitem[{{Corsaro} \& {De Ridder}(2014{\natexlab{a}})}]{corsaro1}
{Corsaro}, E. \& {De Ridder}, J. 2014{\natexlab{a}}, \aap, 571, A71

\bibitem[{{Corsaro} \& {De Ridder}(2014{\natexlab{b}})}]{corsaro2}
{Corsaro}, E. \& {De Ridder}, J. 2014{\natexlab{b}}, {DIAMONDS:
  high-DImensional And multi-MOdal NesteD Sampling}, Astrophysics Source Code
  Library

\bibitem[{Corsaro {et~al.}(2018)Corsaro, De~Ridder, \& García}]{corsaro3}
Corsaro, E., De~Ridder, J., \& García, R.~A. 2018, Astronomy and Astrophysics,
  612, C2

\bibitem[{{Desmars} {et~al.}(2015){Desmars}, {Camargo}, {Braga-Ribas},
  {Vieira-Martins}, {Assafin}, {Vachier}, {Colas}, {Ortiz}, {Duffard},
  {Morales}, {Sicardy}, {Gomes-J{\'u}nior}, \& {Benedetti-Rossi}}]{desmars2013}
{Desmars}, J., {Camargo}, J.~I.~B., {Braga-Ribas}, F., {et~al.} 2015, \aap,
  584, A96

\bibitem[{{Dunham} {et~al.}(2002){Dunham}, {Goffin}, {Manek}, {Federspiel},
  {Stone}, \& {Owen}}]{dunham}
{Dunham}, D.~W., {Goffin}, E., {Manek}, J., {et~al.} 2002, \memsai, 73, 662

\bibitem[{{\v{D}urech} {et~al.}(2015){\v{D}urech}, {Carry}, {Delbo},
  {Kaasalainen}, \& {Viikinkoski}}]{durech}
{\v{D}urech}, J., {Carry}, B., {Delbo}, M., {Kaasalainen}, M., \&
  {Viikinkoski}, M. 2015, {Asteroid Models from Multiple Data Sources}
  ({Michel}, P. and {DeMeo}, F.~E. and {Bottke}, W.~F.), 183--202

\bibitem[{Farnocchia {et~al.}(2015)Farnocchia, Chesley, Chamberlin, \&
  Tholen}]{farnocchia_2015}
Farnocchia, D., Chesley, S.~R., Chamberlin, A.~B., \& Tholen, D.~J. 2015,
  Icarus, 245, 94

\bibitem[{{Gaia Collaboration} {et~al.}(2018{\natexlab{a}}){Gaia
  Collaboration}, {Brown}, {Vallenari}, {Prusti}, {de Bruijne}, {Babusiaux},
  {Bailer-Jones}, {Biermann}, {Evans}, {Eyer}, \& et~al.}]{gaia4}
{Gaia Collaboration}, {Brown}, A.~G.~A., {Vallenari}, A., {et~al.}
  2018{\natexlab{a}}, \aap, 616, A1

\bibitem[{{Gaia Collaboration} {et~al.}(2016){Gaia Collaboration}, {Brown},
  {Vallenari}, {Prusti}, {de Bruijne}, {Mignard}, {Drimmel}, {Babusiaux},
  {Bailer-Jones}, {Bastian}, \& et~al.}]{gaia2}
{Gaia Collaboration}, {Brown}, A.~G.~A., {Vallenari}, A., {et~al.} 2016, \aap,
  595, A2

\bibitem[{{Gaia Collaboration} {et~al.}(2018{\natexlab{b}}){Gaia
  Collaboration}, {Spoto}, {Tanga}, {Mignard}, {Berthier}, {Carry}, {Cellino},
  {Dell'Oro}, {Hestroffer}, {Muinonen}, \& et~al.}]{gaia3}
{Gaia Collaboration}, {Spoto}, F., {Tanga}, P., {et~al.} 2018{\natexlab{b}},
  \aap, 616, A13

\bibitem[{Herald {et~al.}(2019)Herald, Frappa, Gault, Hayamizu, Kerr, Moore, \&
  Giacchini}]{herald_2019}
Herald, D., Frappa, E., Gault, D., {et~al.} 2019, NASA Planetary Data System,
  Asteroid Occultations V3.0. urn:nasa:pds:smallbodiesoccultations::3.0.

\bibitem[{Hunter(2007)}]{Hunter2007}
Hunter, J.~D. 2007, Computing in Science \& Engineering, 9, 90

\bibitem[{Ivezić {et~al.}(2019)Ivezić, Kahn, Tyson, Abel, Acosta, Allsman,
  Alonso, AlSayyad, Anderson, Andrew, Angel, Angeli, Ansari, Antilogus, Araujo,
  Armstrong, Arndt, Astier, Aubourg, Auza, Axelrod, Bard, Barr, Barrau,
  Bartlett, Bauer, Bauman, Baumont, Bechtol, Bechtol, Becker, Becla, Beldica,
  Bellavia, Bianco, Biswas, Blanc, Blazek, Blandford, Bloom, Bogart, Bond,
  Booth, Borgland, Borne, Bosch, Boutigny, Brackett, Bradshaw, Brandt, Brown,
  Bullock, Burchat, Burke, Cagnoli, Calabrese, Callahan, Callen, Carlin,
  Carlson, Chandrasekharan, Charles-Emerson, Chesley, Cheu, Chiang, Chiang,
  Chirino, Chow, Ciardi, Claver, Cohen-Tanugi, Cockrum, Coles, Connolly, Cook,
  Cooray, Covey, Cribbs, Cui, Cutri, Daly, Daniel, Daruich, Daubard, Daues,
  Dawson, Delgado, Dellapenna, de~Peyster, de~Val-Borro, Digel, Doherty,
  Dubois, Dubois-Felsmann, Durech, Economou, Eifler, Eracleous, Emmons,
  Fausti~Neto, Ferguson, Figueroa, Fisher-Levine, Focke, Foss, Frank, Freemon,
  Gangler, Gawiser, Geary, Gee, Geha, Gessner, Gibson, Gilmore, Glanzman,
  Glick, Goldina, Goldstein, Goodenow, Graham, Gressler, Gris, Guy, Guyonnet,
  Haller, Harris, Hascall, Haupt, Hernandez, Herrmann, Hileman, Hoblitt,
  Hodgson, Hogan, Howard, Huang, Huffer, Ingraham, Innes, Jacoby, Jain, Jammes,
  Jee, Jenness, Jernigan, Jevremović, Johns, Johnson, Johnson, Jones,
  Juramy-Gilles, Jurić, Kalirai, Kallivayalil, Kalmbach, Kantor, Karst,
  Kasliwal, Kelly, Kessler, Kinnison, Kirkby, Knox, Kotov, Krabbendam,
  Krughoff, Kubánek, Kuczewski, Kulkarni, Ku, Kurita, Lage, Lambert, Lange,
  Langton, Le~Guillou, Levine, Liang, Lim, Lintott, Long, Lopez, Lotz, Lupton,
  Lust, MacArthur, Mahabal, Mandelbaum, Markiewicz, Marsh, Marshall, Marshall,
  May, McKercher, McQueen, Meyers, Migliore, Miller, Mills, Miraval, Moeyens,
  Moolekamp, Monet, Moniez, Monkewitz, Montgomery, Morrison, Mueller, Muller,
  Muñoz~Arancibia, Neill, Newbry, Nief, Nomerotski, Nordby, O'Connor, Oliver,
  Olivier, Olsen, O'Mullane, Ortiz, Osier, Owen, Pain, Palecek, Parejko,
  Parsons, Pease, Peterson, Peterson, Petravick, Libby~Petrick, Petry,
  Pierfederici, Pietrowicz, Pike, Pinto, Plante, Plate, Plutchak, Price,
  Prouza, Radeka, Rajagopal, Rasmussen, Regnault, Reil, Reiss, Reuter, Ridgway,
  Riot, Ritz, Robinson, Roby, Roodman, Rosing, Roucelle, Rumore, Russo, Saha,
  Sassolas, Schalk, Schellart, Schindler, Schmidt, Schneider, Schneider,
  Schoening, Schumacher, Schwamb, Sebag, Selvy, Sembroski, Seppala, Serio,
  Serrano, Shaw, Shipsey, Sick, Silvestri, Slater, Smith, Smith, Sobhani,
  Soldahl, Storrie-Lombardi, Stover, Strauss, Street, Stubbs, Sullivan,
  Sweeney, Swinbank, Szalay, Takacs, Tether, Thaler, Thayer, Thomas, Thornton,
  Thukral, Tice, Trilling, Turri, Van~Berg, Vanden~Berk, Vetter, Virieux,
  Vucina, Wahl, Walkowicz, Walsh, Walter, Wang, Wang, Warner, Wiecha, Willman,
  Winters, Wittman, Wolff, Wood-Vasey, Wu, Xin, Yoachim, \& Zhan}]{ivezic_2019}
Ivezić, v., Kahn, S.~M., Tyson, J.~A., {et~al.} 2019, The Astrophysical
  Journal, 873, 111

\bibitem[{Mainzer {et~al.}(2011{\natexlab{a}})Mainzer, Grav, Bauer, Masiero,
  McMillan, Cutri, Walker, Wright, Eisenhardt, Tholen, Spahr, Jedicke, Denneau,
  DeBaun, Elsbury, Gautier, Gomillion, Hand, Mo, Watkins, Wilkins, Bryngelson,
  Molina, Desai, Camus, Hidalgo, Konstantopoulos, Larsen, Maleszewski, Malkan,
  Mauduit, Mullan, Olszewski, Pforr, Saro, Scotti, \&
  Wasserman}]{mainzer_2011b}
Mainzer, A., Grav, T., Bauer, J., {et~al.} 2011{\natexlab{a}}, The
  Astrophysical Journal, 743, 156

\bibitem[{Mainzer {et~al.}(2011{\natexlab{b}})Mainzer, Grav, Masiero, Hand,
  Bauer, Tholen, McMillan, Spahr, Cutri, Wright, Watkins, Mo, \&
  Maleszewski}]{mainzer_2011a}
Mainzer, A., Grav, T., Masiero, J., {et~al.} 2011{\natexlab{b}}, The
  Astrophysical Journal, 741, 90

\bibitem[{{Mary}(2006)}]{mary}
{Mary}, D.~L. 2006, \aap, 452, 715

\bibitem[{Masiero {et~al.}(2011)Masiero, Mainzer, Grav, Bauer, Cutri, Dailey,
  Eisenhardt, McMillan, Spahr, Skrutskie, Tholen, Walker, Wright, DeBaun,
  Elsbury, Gautier, Gomillion, \& Wilkins}]{masiero_2011}
Masiero, J.~R., Mainzer, A.~K., Grav, T., {et~al.} 2011, The Astrophysical
  Journal, 741, 68

\bibitem[{{Meza} {et~al.}(2019){Meza}, {Sicardy}, {Assafin}, {Ortiz},
  {Bertrand}, {Lellouch}, {Desmars}, {Forget}, {B{\'e}rard}, {Doressoundiram},
  {Lecacheux}, {Marques Oliveira}, {Roques}, {Widemann}, {Colas}, {Vachier},
  {Renner}, {Leiva}, {Braga-Ribas}, {Benedetti-Rossi}, {Camargo},
  {Dias-Oliveira}, {Morgado}, {Gomes-J{\'u}nior}, {Vieira-Martins}, {Behrend},
  {Castro Tirado}, {Duffard}, {Morales}, {Santos-Sanz}, {Jel{\'{\i}}nek},
  {Cunniffe}, {Querel}, {Harnisch}, {Jansen}, {Pennell}, {Todd}, {Ivanov},
  {Opitom}, {Gillon}, {Jehin}, {Manfroid}, {Pollock}, {Reichart}, {Haislip},
  {Ivarsen}, {LaCluyze}, {Maury}, {Gil-Hutton}, {Dhillon}, {Littlefair},
  {Marsh}, {Veillet}, {Bath}, {Beisker}, {Bode}, {Kretlow}, {Herald}, {Gault},
  {Kerr}, {Pavlov}, {Farag{\'o}}, {Kl{\"o}s}, {Frappa}, {Lavayssi{\`e}re},
  {Cole}, {Giles}, {Greenhill}, {Hill}, {Buie}, {Olkin}, {Young}, {Young},
  {Wasserman}, {Devog{\`e}le}, {French}, {Bianco}, {Marchis}, {Brosch},
  {Kaspi}, {Polishook}, {Manulis}, {Ait Moulay Larbi}, {Benkhaldoun},
  {Daassou}, {El Azhari}, {Moulane}, {Broughton}, {Milner}, {Dobosz}, {Bolt},
  {Lade}, {Gilmore}, {Kilmartin}, {Allen}, {Graham}, {Loader}, {McKay},
  {Talbot}, {Parker}, {Abe}, {Bendjoya}, {Rivet}, {Vernet}, {Di Fabrizio},
  {Lorenzi}, {Magazz{\`u}}, {Molinari}, {Gazeas}, {Tzouganatos}, {Carbognani},
  {Bonnoli}, {Marchini}, {Leto}, {Zanmar Sanchez}, {Mancini}, {Kattentidt},
  {Dohrmann}, {Guhl}, {Rothe}, {Walzel}, {Wortmann}, {Eberle}, {Hampf},
  {Ohlert}, {Krannich}, {Murawsky}, {G{\"a}hrken}, {Gloistein}, {Alonso},
  {Rom{\'a}n}, {Communal}, {Jabet}, {de Visscher}, {S{\'e}rot}, {Janik},
  {Moravec}, {Machado}, {Selva}, {Perell{\'o}}, {Rovira}, {Conti}, {Papini},
  {Salvaggio}, {Noschese}, {Tsamis}, {Tigani}, {Barroy}, {Irzyk}, {Neel},
  {Godard}, {Lanoisel{\'e}e}, {Sogorb}, {V{\'e}rilhac}, {Bretton}, {Signoret},
  {Ciabattari}, {Naves}, {Boutet}, {De Queiroz}, {Lindner}, {Lindner},
  {Enskonatus}, {Dangl}, {Tordai}, {Eichler}, {Hattenbach}, {Peterson},
  {Molnar}, \& {Howell}}]{pluto2}
{Meza}, E., {Sicardy}, B., {Assafin}, M., {et~al.} 2019, arXiv e-prints
  [\eprint[arXiv]{1903.02315}]

\bibitem[{{Ortiz} {et~al.}(2012){Ortiz}, {Sicardy}, {Braga-Ribas},
  {Alvarez-Candal}, {Lellouch}, {Duffard}, {Pinilla-Alonso}, {Ivanov},
  {Littlefair}, {Camargo}, {Assafin}, {Unda-Sanzana}, {Jehin}, {Morales},
  {Tancredi}, {Gil-Hutton}, {de La Cueva}, {Colque}, {da Silva Neto},
  {Manfroid}, {Thirouin}, {Guti{\'e}rrez}, {Lecacheux}, {Gillon}, {Maury},
  {Colas}, {Licandro}, {Mueller}, {Jacques}, {Weaver}, {Milone}, {Salvo},
  {Bruzzone}, {Organero}, {Behrend}, {Roland}, {Vieira-Martins}, {Widemann},
  {Roques}, {Santos-Sanz}, {Hestroffer}, {Dhillon}, {Marsh}, {Harlingten},
  {Campo Bagatin}, {Alonso}, {Ortiz}, {Colazo}, {Lima}, {Oliveira}, {Kerber},
  {Smiljanic}, {Pimentel}, {Giacchini}, {Cacella}, \& {Emilio}}]{makemake}
{Ortiz}, J.~L., {Sicardy}, B., {Braga-Ribas}, F., {et~al.} 2012, \nat, 491, 566

\bibitem[{{Ortiz} {et~al.}(2019){Ortiz}, {Sicardy}, {Camargo}, {Santos-Sanz},
  \& {Braga-Ribas}}]{Ortiz_2019}
{Ortiz}, J.~L., {Sicardy}, B., {Camargo}, J.~I.~B., {Santos-Sanz}, P., \&
  {Braga-Ribas}, F. 2019, arXiv e-prints, arXiv:1905.04335

\bibitem[{Ostro {et~al.}(2002)Ostro, Hudson, Benner, Giorgini, Magri, Margot,
  \& Nolan}]{ostro_2002}
Ostro, S.~J., Hudson, R.~S., Benner, L. A.~M., {et~al.} 2002, Asteroids III,
  151

\bibitem[{{Pass} {et~al.}(2018){Pass}, {Metchev}, {Brown}, \&
  {Beauchemin}}]{pass}
{Pass}, E., {Metchev}, S., {Brown}, P., \& {Beauchemin}, S. 2018, \pasp, 130,
  014502

\bibitem[{{Perryman}(2012)}]{perryman}
{Perryman}, M. 2012, European Physical Journal H, 37, 745

\bibitem[{{Porter} {et~al.}(2018){Porter}, {Buie}, {Parker}, {Spencer},
  {Benecchi}, {Tanga}, {Verbiscer}, {Kavelaars}, {Gwyn}, {Young}, {Weaver},
  {Olkin}, {Parker}, \& {Stern}}]{Porter_2018}
{Porter}, S.~B., {Buie}, M.~W., {Parker}, A.~H., {et~al.} 2018, \aj, 156, 20

\bibitem[{Pravec {et~al.}(2012)Pravec, Harris, Kušnirák, Galád, \&
  Hornoch}]{pravec_2012}
Pravec, P., Harris, A.~W., Kušnirák, P., Galád, A., \& Hornoch, K. 2012,
  Icarus, 221, 365

\bibitem[{{Price-Whelan} {et~al.}(2018){Price-Whelan}, {Sip{\H{o}}cz},
  {G{\"u}nther}, {Lim}, {Crawford}, {Conseil}, {Shupe}, {Craig}, {Dencheva},
  {Ginsburg}, {VanderPlas}, {Bradley}, {P{\'e}rez-Su{\'a}rez}, {de Val-Borro},
  {Paper Contributors}, {Aldcroft}, {Cruz}, {Robitaille}, {Tollerud},
  {Coordination Committee}, {Ardelean}, {Babej}, {Bach}, {Bachetti}, {Bakanov},
  {Bamford}, {Barentsen}, {Barmby}, {Baumbach}, {Berry}, {Biscani}, {Boquien},
  {Bostroem}, {Bouma}, {Brammer}, {Bray}, {Breytenbach}, {Buddelmeijer},
  {Burke}, {Calderone}, {Cano Rodr{\'\i}guez}, {Cara}, {Cardoso}, {Cheedella},
  {Copin}, {Corrales}, {Crichton}, {D{\textquoteright}Avella}, {Deil},
  {Depagne}, {Dietrich}, {Donath}, {Droettboom}, {Earl}, {Erben}, {Fabbro},
  {Ferreira}, {Finethy}, {Fox}, {Garrison}, {Gibbons}, {Goldstein}, {Gommers},
  {Greco}, {Greenfield}, {Groener}, {Grollier}, {Hagen}, {Hirst}, {Homeier},
  {Horton}, {Hosseinzadeh}, {Hu}, {Hunkeler}, {Ivezi{\'c}}, {Jain}, {Jenness},
  {Kanarek}, {Kendrew}, {Kern}, {Kerzendorf}, {Khvalko}, {King}, {Kirkby},
  {Kulkarni}, {Kumar}, {Lee}, {Lenz}, {Littlefair}, {Ma}, {Macleod},
  {Mastropietro}, {McCully}, {Montagnac}, {Morris}, {Mueller}, {Mumford},
  {Muna}, {Murphy}, {Nelson}, {Nguyen}, {Ninan}, {N{\"o}the}, {Ogaz}, {Oh},
  {Parejko}, {Parley}, {Pascual}, {Patil}, {Patil}, {Plunkett}, {Prochaska},
  {Rastogi}, {Reddy Janga}, {Sabater}, {Sakurikar}, {Seifert}, {Sherbert},
  {Sherwood-Taylor}, {Shih}, {Sick}, {Silbiger}, {Singanamalla}, {Singer},
  {Sladen}, {Sooley}, {Sornarajah}, {Streicher}, {Teuben}, {Thomas},
  {Tremblay}, {Turner}, {Terr{\'o}n}, {van Kerkwijk}, {de la Vega}, {Watkins},
  {Weaver}, {Whitmore}, {Woillez}, {Zabalza}, \& {Contributors}}]{astropy:2018}
{Price-Whelan}, A.~M., {Sip{\H{o}}cz}, B.~M., {G{\"u}nther}, H.~M., {et~al.}
  2018, \aj, 156, 123

\bibitem[{{Roques} {et~al.}(2008){Roques}, {Georgevits}, \&
  {Doressoundiram}}]{roques}
{Roques}, F., {Georgevits}, G., \& {Doressoundiram}, A. 2008, {The Kuiper Belt
  Explored by Serendipitous Stellar Occultations} ({Barucci}, M.~A. and
  {Boehnhardt}, H. and {Cruikshank}, D.~P. and {Morbidelli}, A. and {Dotson},
  R.), 545--556

\bibitem[{Scheeres {et~al.}(2015)Scheeres, Britt, Carry, \&
  Holsapple}]{scheeres_2015}
Scheeres, D.~J., Britt, D., Carry, B., \& Holsapple, K.~A. 2015, Asteroids IV,
  745

\bibitem[{{Shevchenko} \& {Tedesco}(2006)}]{shev}
{Shevchenko}, V.~G. \& {Tedesco}, E.~F. 2006, \icarus, 184, 211

\bibitem[{{Sicardy} {et~al.}(2011){Sicardy}, {Ortiz}, {Assafin}, {Jehin},
  {Maury}, {Lellouch}, {Hutton}, {Braga-Ribas}, {Colas}, {Hestroffer},
  {Lecacheux}, {Roques}, {Santos-Sanz}, {Widemann}, {Morales}, {Duffard},
  {Thirouin}, {Castro-Tirado}, {Jel{\'{\i}}nek}, {Kub{\'a}nek}, {Sota},
  {S{\'a}nchez-Ram{\'{\i}}rez}, {Andrei}, {Camargo}, {da Silva Neto}, {Gomes},
  {Martins}, {Gillon}, {Manfroid}, {Tozzi}, {Harlingten}, {Saravia}, {Behrend},
  {Mottola}, {Melendo}, {Peris}, {Fabregat}, {Madiedo}, {Cuesta}, {Eibe},
  {Ull{\'a}n}, {Organero}, {Pastor}, {de Los Reyes}, {Pedraz}, {Castro}, {de La
  Cueva}, {Muler}, {Steele}, {Cebri{\'a}n},
  {Monta{\~n}{\'e}s-Rodr{\'{\i}}guez}, {Oscoz}, {Weaver}, {Jacques}, {Corradi},
  {Santos}, {Reis}, {Milone}, {Emilio}, {Guti{\'e}rrez}, {V{\'a}zquez}, \&
  {Hern{\'a}ndez-Toledo}}]{eris}
{Sicardy}, B., {Ortiz}, J.~L., {Assafin}, M., {et~al.} 2011, \nat, 478, 493

\bibitem[{{Sicardy} {et~al.}(2003){Sicardy}, {Widemann}, {Lellouch}, {Veillet},
  {Cuillandre}, {Colas}, {Roques}, {Beisker}, {Kretlow}, {Lagrange}, {Gendron},
  {Lacombe}, {Lecacheux}, {Birnbaum}, {Fienga}, {Leyrat}, {Maury}, {Raynaud},
  {Renner}, {Schultheis}, {Brooks}, {Delsanti}, {Hainaut}, {Gilmozzi},
  {Lidman}, {Spyromilio}, {Rapaport}, {Rosenzweig}, {Naranjo}, {Porras},
  {D{\'{\i}}az}, {Calder{\'o}n}, {Carrillo}, {Carvajal}, {Recalde}, {Cavero},
  {Montalvo}, {Barr{\'{\i}}a}, {Campos}, {Duffard}, \& {Levato}}]{pluto}
{Sicardy}, B., {Widemann}, T., {Lellouch}, E., {et~al.} 2003, \nat, 424, 168

\bibitem[{Spoto {et~al.}(2015)Spoto, Milani, \& Knezevic}]{spoto_2015}
Spoto, F., Milani, A., \& Knezevic, Z. 2015, arXiv:1504.05461 [astro-ph],
  arXiv: 1504.05461

\bibitem[{{Spoto} {et~al.}(2017){Spoto}, {Tanga}, {Bouquillon}, {Desmars},
  {Hestroffer}, {Mignard}, {Altmann}, {Herald}, {Marchant}, {Barache},
  {Carlucci}, {Lister}, \& {Taris}}]{spoto}
{Spoto}, F., {Tanga}, P., {Bouquillon}, S., {et~al.} 2017, \aap, 607, A21

\bibitem[{{Tanga} \& {Delbo}(2007)}]{tanga1}
{Tanga}, P. \& {Delbo}, M. 2007, \aap, 474, 1015

\bibitem[{{Virtanen} {et~al.}(2019){Virtanen}, {Gommers}, {Oliphant},
  {Haberland}, {Reddy}, {Cournapeau}, {Burovski}, {Peterson}, {Weckesser},
  {Bright}, {van der Walt}, {Brett}, {Wilson}, {Jarrod Millman}, {Mayorov},
  {Nelson}, {Jones}, {Kern}, {Larson}, {Carey}, {Polat}, {Feng}, {Moore}, {Vand
  erPlas}, {Laxalde}, {Perktold}, {Cimrman}, {Henriksen}, {Quintero}, {Harris},
  {Archibald}, {Ribeiro}, {Pedregosa}, {van Mulbregt}, \&
  {Contributors}}]{scipy2019}
{Virtanen}, P., {Gommers}, R., {Oliphant}, T.~E., {et~al.} 2019, arXiv
  e-prints, arXiv:1907.10121

\bibitem[{Vokrouhlický {et~al.}(2015)Vokrouhlický, Bottke, Chesley, Scheeres,
  \& Statler}]{vokro_2015}
Vokrouhlický, D., Bottke, W.~F., Chesley, S.~R., Scheeres, D.~J., \& Statler,
  T.~S. 2015, The {Yarkovsky} and {YORP} {Effects} (University of Arizona
  Press), 509--531

\end{thebibliography}
